\documentclass[twocolumn]{aastex631}

\usepackage{amsmath}

\shorttitle{Ambipolar electric field and potential in the solar wind}
\shortauthors{Ber\v{c}i\v{c} et al.}
\graphicspath{{./}{figures/}}

\begin{document}

\title{Ambipolar electric field and potential in the solar wind estimated from electron velocity distribution functions}

\newcommand{\sao}{\affiliation{Smithsonian Astrophysical Observatory, Cambridge, MA, USA}}
\newcommand{\umich}{\affiliation{University of Michigan, Ann Arbor, MI, USA}}
\newcommand{\ucb}{\affiliation{Physics Department,University of California, Berkeley, CA, USA}}
\newcommand{\les}{\affiliation{LESIA, Observatoire de Paris, Universit\'e PSL, CNRS, Sorbonne Universit\'e, Unversit\'e de Paris, 5 place Jules Janssen, 92195 Meudon, France}}
\newcommand{\ufi}{\affiliation{Physics and Astronomy Department, University of Florence, Sesto Fiorentino, Italy}}
\newcommand{\ina}{\affiliation{INAF - Osservatorio Astrofisico di Arcetri, Firenze, Italy}}
\newcommand{\ssl}{\affiliation{Space Sciences Laboratory, University of California, Berkeley, CA 94720-7450, USA}}

\newcommand{\mssl}{\affiliation{Mullard Space Science Laboratory, University College London, Dorking, RH5 6NT, UK}}

\newcommand{\imp}{\affiliation{The Blackett Laboratory, Imperial College London, London, SW7 2AZ, UK}}
\newcommand{\que}{\affiliation{School of Physics and Astronomy, Queen Mary University of London, London E1 4NS, UK}}

\newcommand{\orl}{\affiliation{LPC2E, CNRS and University of Orl\'eans, Orl\'eans, France}}

\newcommand{\mine}{\affiliation{School of Physics and Astronomy, University of Minnesota, Minneapolis, MN 55455, USA}}

\newcommand{\nas}{\affiliation{Solar System Exploration Division, NASA/Goddard Space Flight Center, Greenbelt, MD, 20771, USA}}

\newcommand{\lasp}{\affiliation{Astrophysical and Planetary Sciences Department, University of Colorado, Boulder, CO, USA}}

\newcommand{\uio}{\affiliation{Department of Physics and Astronomy, University of Iowa, Iowa City, IA 52242, USA}}
\newcommand{\unh}{\affiliation{Space Science Center, University of New Hampshire, 8 College Road, Durham NH 03824, USA}}

\correspondingauthor{Laura Ber\v{c}i\v{c}}
\email{l.bercic@ucl.ac.uk}
\author[0000-0002-6075-1813]{Laura Ber\v{c}i\v{c}}\mssl %
\author[0000-0001-6172-5062]{Milan Maksimovi\'{c}}\les  %
\author[0000-0001-5258-6128]{Jasper S. Halekas}\uio     %
\author[0000-0002-1647-6121]{Simone Landi}\ufi\ina      %
\author[0000-0003-0293-8601]{Christopher J. Owen}\mssl  %
\author[0000-0002-0497-1096]{Daniel Verscharen}\mssl \unh %
\author[0000-0001-5030-6030]{Davin Larson}\ucb          %
\author[0000-0002-7287-5098]{Phyllis Whittlesey}\ucb    %
\author[0000-0002-6145-436X]{Samuel T. Badman}\ucb      %

\author[0000-0002-1989-3596]{Stuart. D. Bale}\ucb\ssl   
\author[0000-0002-3520-4041]{Anthony W. Case}\sao       
\author[0000-0003-0420-3633]{Keith Goetz}\mine          
\author[0000-0002-6938-0166]{Peter R. Harvey}\ssl       
\author[0000-0002-7077-930X]{Justin C. Kasper}\umich\sao
\author[0000-0001-6095-2490]{Kelly E. Korreck}\sao      
\author[0000-0002-0396-0547]{Roberto Livi}\ucb          
\author[0000-0003-3112-4201]{Robert J. MacDowall}\nas   
\author[0000-0003-1191-1558]{David M. Malaspina}\lasp   %
\author[0000-0002-1573-7457]{Marc Pulupa}\ssl
\author[0000-0002-7728-0085]{Michael L. Stevens}\sao

 
\begin{abstract}
The solar wind escapes from the solar corona and is accelerated, over a short distance, to its terminal velocity. The energy balance associated with this acceleration remains poorly understood.
To quantify the global electrostatic contribution to the solar wind dynamics, we empirically estimate the ambipolar electric field ($\mathrm{E}_\parallel$) and potential ($\Phi_\mathrm{r,\infty}$).
We analyse electron velocity distribution functions (VDFs) measured in the near-Sun solar wind, between 20.3\,$R_S$ and 85.3\,$R_S$, by the Parker Solar Probe. We test the predictions of two different solar wind models.
Close to the Sun, the VDFs exhibit a suprathermal electron deficit in the sunward, magnetic field aligned part of phase space. We argue that the sunward deficit is a remnant of the electron cutoff predicted by  collisionless exospheric models \citep{Lemaire1970,LemaireJosephandScherer1971, Jockers1970}. This cutoff energy is directly linked to $\Phi_\mathrm{r,\infty}$.
Competing effects of $\mathrm{E}_\parallel$ and Coulomb collisions in the solar wind are addressed by the Steady Electron Runaway Model (SERM) \citep{Scudder2019serm}. In this model, electron phase space is separated into collisionally overdamped and underdamped regions. We assume that this boundary velocity at small pitch angles coincides with the strahl break-point energy, which allows us to calculate $\mathrm{E}_\parallel$.
The obtained $\Phi_\mathrm{r,\infty}$ and $\mathrm{E}_\parallel$ agree well with theoretical expectations. They decrease with radial distance as power law functions with indices $\alpha_\Phi = -0.66$ and $\alpha_\mathrm{E} = -1.69$. We finally estimate the velocity gained by protons from electrostatic acceleration, which equals to 77\% calculated from the exospheric models, and to 44\% from the SERM model.

\end{abstract}



\section{Introduction} \label{sec:intro}

The solar wind is a continuous outflow of plasma from the hot solar corona \citep{Parker1958}. The particles escaping the Sun are mostly electrons and protons, with a smaller population of heavier ions. Over a small radial distance, these particles reach bulk velocities of order a few 100\,km\,s$^{-1}$. The nature of the acceleration mechanisms converting the coronal thermal energy to the solar wind kinetic energy remains one of the most important open questions in heliophysics.

The terminal velocity of the solar wind is closely related to the density and temperature of the solar coronal plasma. These can be estimated remotely through spectroscopy and multi-frequency radio imaging \citep{Marcier2015}. In coronal holes, which are regions of open magnetic field lines along which plasma can flow freely in the radial direction, the typical electron temperature is 0.79 MK \citep{David1998, Cranmer2002}, while much higher temperatures are found on the edges of coronal holes and in active regions \citep{Stansby2020}.

In coronal holes, the proton distributions appear hotter than that of electrons, and anisotropic with $T_\perp > T_\parallel$ at radial distance $\sim$ 3\,$R_S$ \citep{Cranmer2002}. Heavier ion distributions are strongly anisotropic at these distances with $T_\perp / T_\parallel$ ranging between 10 and 100 \citep{Kohl1998}. Preferential perpendicular ion heating is believed to contribute to the solar wind acceleration \citep{MunroJackson1977} through mechanisms like stochastic heating \citep{Chandran2010, Bourouaine2013}, ion-cyclotron resonance \citep{Dusenbery1981, Hollweg1999, Li1999, MarschTu2001, TuMarsch2001} or the dissipation of turbulence \citep{Bieber1996, Oughton2001,verdini2010, Karimabadi2013, Matthaeus2015, AgudeloRueda2021}.

In the case of electrons, a non-Maxwellian coronal VDF with an excess of high-energy electrons can alone accelerate the solar wind protons to velocities above 700\,km\,s$^{-1}$ \citep{Maksimovic1997c, Zouganelis2004a}. The radial evolution of the collisionless, expanding solar wind is captured by the exospheric solar wind models \citep{Lemaire1970, LemaireJosephandScherer1971, Jockers1970, Pierrard1999, Maksimovic2001a, Zouganelis2005}. The drivers of the solar wind in these models are the solar wind electrons. Due to their small mass, their thermal velocity just above the solar surface is large enough for the majority of electrons to escape the Sun's gravity. However, this is not the case for the heavier protons. A global electric polarization field, also referred to as the \emph{ambipolar} electric field ($\mathrm{E}_\parallel$), builds up, accelerating the protons and decelerating the electrons. It preserves equal ambipolar diffusion of ions and electrons in the radial direction. This study focuses on the quantification of $\mathrm{E}_\parallel$ in the solar wind and thus its contribution to the total solar wind acceleration.

$\mathrm{E}_\parallel$ decreases with radial distance from the Sun, and has a magnitude of order a few \,nV/m in the inner heliosphere \citep[e.g.][]{Bercic2021}, thus it is practically undetectable by spacecraft electric field antennas. However, electron VDFs measured in the near-Sun solar wind are highly affected by $\mathrm{E}_\parallel$ and thus can tell us something about its properties.

Electron VDFs in the solar wind have a complex structure and are commonly modelled with three components. Low energy electrons belong to the almost isotropic \emph{core} population and are well represented by a Maxwellian distribution. Higher energy electrons belong to either the isotropic \emph{halo} population, or the magnetic field aligned, beam-like population, called the \emph{strahl} \citep{Feldman1975a, SchwartcMarsch1983, Pilipp1987b, Maksimovic1997b, Maksimovic2005a, stverak2008, Stverak2009a, Tao2016}. Another electron feature is often observed in the near-Sun solar wind -- a relative deficit of electrons compared to the Maxwellian core model appears in the suprathermal energy range in the portion of phase space, opposite to the strahl direction \citep{Halekas2019, Halekas2020arxiv, Bercic2020, Bercic2021wh, Bercic2021}. The statistical properties of this \emph{sunward deficit} are presented by \citet{Halekas2021submitted}.

\subsection{The exospheric prediction}

A deficit of sunward moving electrons is also a feature of  collisionless exospheric models, where it is referred to as the "electron cutoff". The electron VDF at any radial distance in these models is separated into two parts: anti-sunward moving electrons with energy greater than the ambipolar electric potential energy ($\mathcal{E}_\Phi$) represent the escaping electrons, which focus towards the magnetic field direction and form the strahl population; ballistic electrons with energies less than $\mathcal{E}_\Phi$ represent the core population. In the sunward direction, these electrons are limited by the cutoff energy ($E_\mathrm{C}$) corresponding to the ambipolar potential between their location and the asymptotic value at large heliocentric distances, where $r \rightarrow \infty$  \citep{Jockers1970, Maksimovic2001a}:
\begin{equation}
    \Phi_\mathrm{r,\infty} = E_\mathrm{C} / \mathrm{e}, 
    \label{eq:phi}
\end{equation}
where e is the electron charge and $E_\mathrm{C}$ is defined in the Sun's rest frame. In this paper, we identify $E_\mathrm{C}$ related to the sunward deficit in the electron VDFs observed by Parker Solar Probe (PSP), and use the exospheric prediction to estimate $\Phi_\mathrm{r,\infty}$.

\subsection{The electron runaway model prediction}

A different theoretical description of the solar wind electrons is proposed by \citet{Scudder1996, Scudder2019serm, Scudder2019crisis, Scudder2019astro} termed the Steady Electron Runaway Model (SERM). SERM accounts for the behaviour of weakly-collisional electrons in the large scale ambipolar electric field. Instead of assuming local thermodynamic equilibrium, it proposes a steady electron runaway effect. The Dreicer electric field ($E_\mathrm D$) \citep{Dreicer1959, Dreicer1960} is used to compare the strength of $\mathrm{E}_\parallel$ to the collisionality of the system. $E_\mathrm{D}$ is defined as the constant electric field strength needed to accelerate a thermal particle to twice its velocity in one collision time \citep{Dreicer1959, Dreicer1960}:

\begin{equation}
    E_\mathrm{D} = \frac{2 k_\mathrm{B} T_\mathrm{c\parallel}}{e \lambda_\mathrm{mfp}}, 
    \label{eq:ED}
\end{equation}
where $k_B$ is the Boltzmann constant, $T_\mathrm{c\parallel}$ is the core electron temperature parallel to the magnetic field and $\lambda_\mathrm{mfp}$ is the collisional electron mean-free path.

The resulting electron VDFs consist of collisionally overdamped and underdamped regions, separated by a 2D separatrix in phase space \citep{Fuchs1986}. The overdamped region corresponds to the core population, and the underdamped region to the suprathermal populations. The boundary for small pitch angles can be related to the energy at which the core transitions to the strahl, the strahl break-point energy ($E_\mathrm{BP}$), which we also identify in the PSP electron VDFs. $\mathrm{E}_\parallel$ then follows from \citep{Scudder2019serm}

\begin{equation}
    \mathrm{E}_\parallel = \frac{\alpha k_\mathrm{B} T_\mathrm{c\parallel}}{E_\mathrm{BP}} E_\mathrm{D}, 
    \label{eq:E}
\end{equation}
where $\alpha = 3$ \citep{Dreicer1960} and $E_\mathrm{BP}$ is defined in the plasma frame.\\

In Sec. \ref{sec:data_analysis} we describe the data set and the method we use to obtain $\Phi_\mathrm{r,\infty}$ and $\mathrm{E}_\parallel$. Sec. \ref{sec:results} presents our results, which we discuss in Sec. \ref{sec:discussion}. We summarise our findings and draw conclusions in Sec. \ref{sec:concl}.

\section{Data Analysis} \label{sec:data_analysis}
\subsection{Data Set} \label{sec:data_set}

This work is based on the analysis of the electron VDFs measured by PSP \citep{Fox2016a}, a heliospheric mission exploring the young solar wind near the Sun. The SPAN Electron (SPAN-E) instrument \citep{Whittlesey2020}, part of the SWEAP investigation package on PSP \citep{Kasper2016}, measures the solar wind electrons. SPAN-E consists of two toroidal top-hat analysers, SPAN-A and SPAN-B, which together cover almost a full sky field of view (FOV). A small portion of the combined FOV in the direction of the Sun is blocked by the spacecraft's heat shield, which protects the payload from direct solar radiation. During encounter periods this FOV obstruction affects measurements taken  within $\sim$ 10$^\circ$ from the radial direction. 
Each of the top-hat analysers measures electron velocity directions with 8 small (6$^\circ$) and 8 large (24$^\circ$) azimuth anodes, and 16 elevation deflection states, which vary in angular width from 10$^\circ$ to 15$^\circ$. The electron energy is sampled in 32 log-spaced bins covering the energy range between 2 eV and 2 keV with a $\Delta E/E$ of 7\%. The duration of a full 3D sweep over all energy and deflection bins is 0.218 s. 

We use electron VDFs collected in  Survey Mode during encounter periods, which consist of multiple full 3D sweeps integrated over time. The presented data was collected during PSP's perihelion passages 4 (Jan 23 2020 - Feb 3 2020) and 5 (May 30 2020 - June 15 2020), with the closest approach at 27.8\,$R_S$, and 6 (Sep 16 2020 - Oct 5 2020) and 7 (Jan 10 2021 - Jan 28 2021), with the closest approach at 20.3\,$R_S$. We thus investigate the region between 20.3\,$R_S$ (0.10\,au) and 85.3\,$R_S$ (0.40\,au). For encounters 4, 5 and 7 the integration time is set to 13.95\,s, and for encounter 6 it is set to 3.49\,s. Detailed descriptions of the SPAN-E instruments and their operating modes are provided by \citet{Whittlesey2020}.

In our analysis we also use the magnetic field vector measured by the triaxial fluxgate magnetometer (MAG) part of the FIELDS investigation suite \citep{Bale2016}, and the proton velocity moment derived from the proton VDFs detected by the SPAN Ion (SPAN-I) instrument \citep{Kasper2016}. Both magnetic field and proton velocity are available with higher or equal cadence than the electron VDFs and are thus averaged over the SPAN-E integration periods to match the electron data.

Our main reason for using the data from four orbits of PSP, out of eight in total to date, is the availability of the SPAN-I data, which improves significantly after encounter 3. A Faraday cup instrument (SPC) \citep{Case2019} also measures solar wind proton velocity on PSP, but discrepancies between the two instruments exist \citep{Woodham2021}. We choose to use SPAN-I data which provides more accurate data closer to the Sun, where the aberration allows the solar wind protons to fly into the instrument protected by the heat shield. Other reasons for the data selection are the changes made in the integration time and elevation deflection tables over the course of the first three orbits, ensuring optimal operation of SPAN-E during the later encounters. We also exclude the electron VDFs obtained at larger distances from the Sun during PSP cruise phase, which are integrated over larger time intervals. Our goal is to obtain a consistent data set of electron VDFs and use it to investigate features typical for the near-Sun environment.

\subsection{Method} \label{sec:method}

\begin{figure}
\plotone{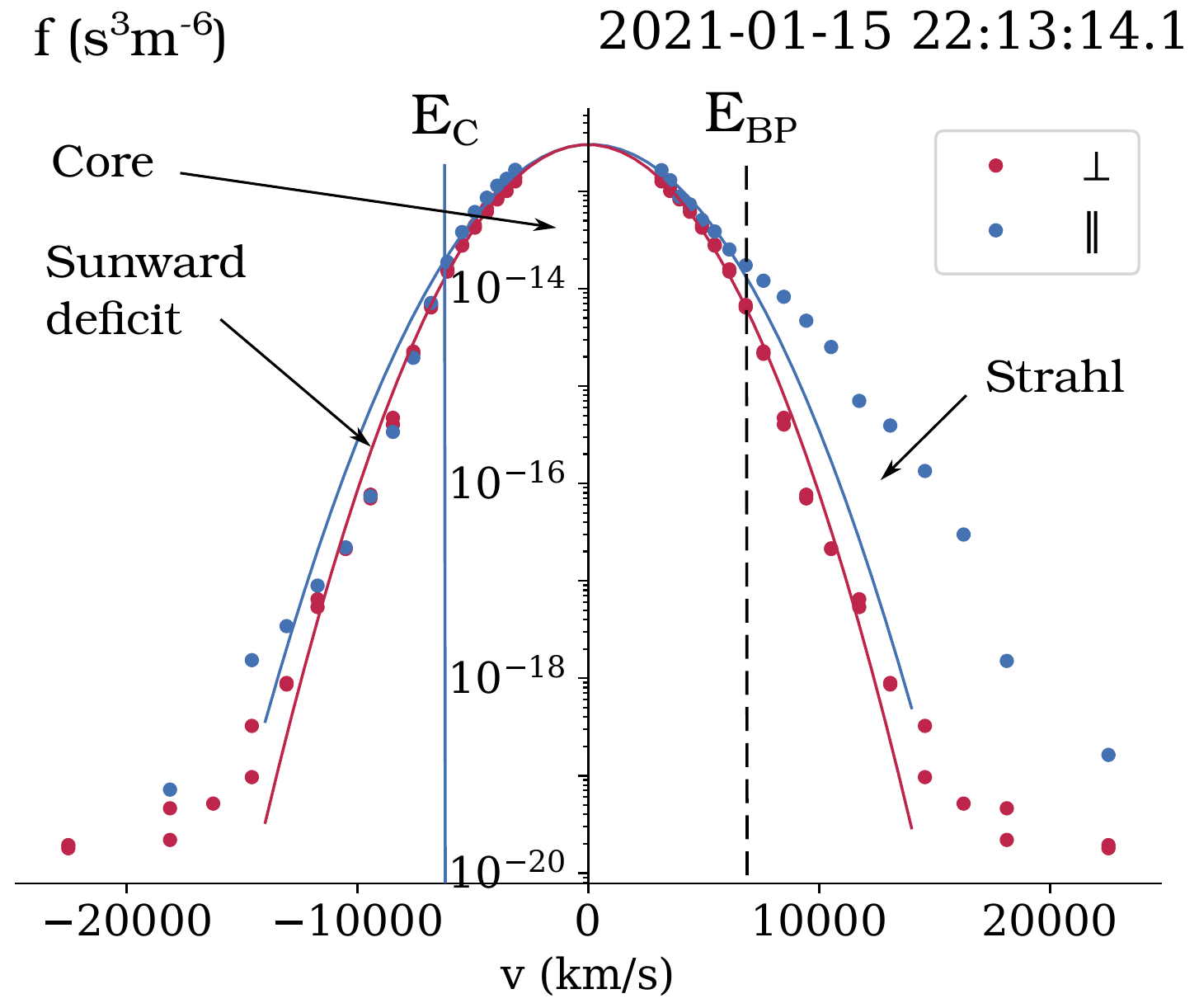}
    \caption{Parallel and perpendicular cuts through an electron VDF in the magnetic field aligned, instrument centred frame. Blue and red dots are the measured points, while blue and red lines denote the parallel and perpendicular cuts through the 3D bi-Maxwellian fit to the core electron population. The blue and black dashed lines mark the transitions between the suprathermal deficit and the core ($E_\mathrm C$), and between the strahl and the core ($E_\mathrm{BP}$) in the parallel direction.}
    \label{fig:example_cut}
\end{figure}

Following the example of previous studies \citep[e.g.][]{Bercic2019, Halekas2019, Bercic2020} we analyse electron VDFs in the magnetic-field aligned, plasma rest frame. We rotate the VDFs given in the SPAN-A and SPAN-B instrument frames, using the magnetic field vector, and shift them using the spacecraft and the solar wind velocities. We show an example of an observed VDF in Fig. \ref{fig:example_cut} as cuts along and perpendicular to the magnetic field direction. We already note the features investigated in this work along the parallel direction: the suprathermal deficit and the strahl. Our aim is to identify the energies at which the electron VDF starts to depart from the Maxwellian core in the directions parallel and anti-parallel to the magnetic field.

We fit the core with a bi-Maxwellian distribution function:
\begin{multline}
    f_\mathrm c (v_\mathrm{\perp1},v_\mathrm{\perp2}, v_\parallel) =  
 A_\mathrm c \exp \left(- \frac{(v_\mathrm{\perp1}- v_\mathrm{c\perp1})^2}{w_\mathrm{c \perp}^2} \right.\\
 \left.
- \frac{(v_\mathrm{\perp2}- v_\mathrm{c\perp2})^2}{w_\mathrm{c \perp}^2} - \frac{(v_\parallel- v_\mathrm{c\parallel})^2}{w_\mathrm{c \parallel}^2} \right),
\end{multline}
where  $A_\mathrm c$ is the normalisation factor, $w_\mathrm{c\perp}$ is the perpendicular core thermal velocity, $w_\mathrm{c\parallel}$ is the parallel core thermal velocity, $v_\mathrm{c\perp1,2}$ are the two core perpendicular drift velocities, and $v_\mathrm{c\parallel}$ is the core parallel drift velocity. These quantities are our fit parameters, from which we obtain the core density as

\begin{equation}
n_\mathrm c = A_\mathrm c \pi^{3/2} w_\mathrm{c\perp}^2 w_\mathrm{c\parallel}
\end{equation}
and the core parallel and perpendicular temperatures as
\begin{equation}
    T_\mathrm{c\parallel} = \frac{m_e w_\mathrm{c\parallel}^2}{2k_B}\quad \text{and}\quad  T_\mathrm{c \perp} = \frac{m_e w_\mathrm{c\perp}^2}{2k_B},
\end{equation}
where $m_e$ is the mass of an electron.

Even though $v_\mathrm{c\perp1,2}$ are expected to be 0 \citep[][e.g.]{Pilipp1987b}, we allow for perpendicular core drifts to correct for possible errors in the measured solar wind velocity vector. $v_\mathrm{c\perp1,2}$ we obtain are small, and the resulting fit parameters are not strongly affected by it.

For the core fit, we only use the data points belonging to the core population, which we determine according to the electron energy and pitch angle. Firstly, we avoid the inclusion of photo-electrons and secondary electrons reflected from the spacecraft by setting a lower energy limit to 35.7 eV\footnote{The same method is used by \citet{Bercic2020}; however, \citet{Halekas2019} fit and subtract the secondary electron peak. Both methods  produce  similar results.}.

Then, we avoid the inclusion of the strahl population through a two-step fitting procedure, which is based on the expected strahl break-point energy ($E_\mathrm{BP}$) following from the kinetic solar wind model BiCoP (Binary Collisions in Plasmas) \citet{Bercic2021}. We use this technique because we find that the core fit along the parallel direction is very sensitive to the selection of data points at small pitch angles. Assuming that $E_\mathrm{BP}$ coincides with the separatrix between the collisionally overdamped and underdamped regions \citep{Scudder2019serm}, $E_\mathrm{BP}$ is related to  $T_\mathrm{c\parallel}$ as
\begin{equation}
    E_\mathrm{BP} = \frac{E_\mathrm D}{\mathrm{E}_\parallel} 3 k_\mathrm{B} T_\mathrm{c\parallel}.
\end{equation}

We perform the first fit to all  measurements with pitch angles greater than 60$^\circ$ and energies less than 132\,eV. The parameters related to the first fit are marked with a tilde. $\widetilde{T}_\mathrm{c,\perp}$ obtained from this first fit is already very accurate as the strahl is narrow near the Sun and mainly affects the core electron fit along the parallel direction. To obtain a zero-order estimation of $\widetilde{E}_\mathrm{BP}$, we assume $\mathrm{E}_\parallel \approx E_\mathrm D $ and $\widetilde{T}_\mathrm{c,\perp} \approx \widetilde{T}_\mathrm{c\parallel}$, and calculate $\widetilde{E}_\mathrm{BP}$ as $\widetilde{E}_\mathrm{BP} = 3 k_\mathrm{B} \widetilde{T}_\mathrm{c\perp}$. This energy is then used for the second (final) fit as an upper energy limit for the data points with pitch angles less than 60$^\circ$. 

\begin{figure}
    \plotone{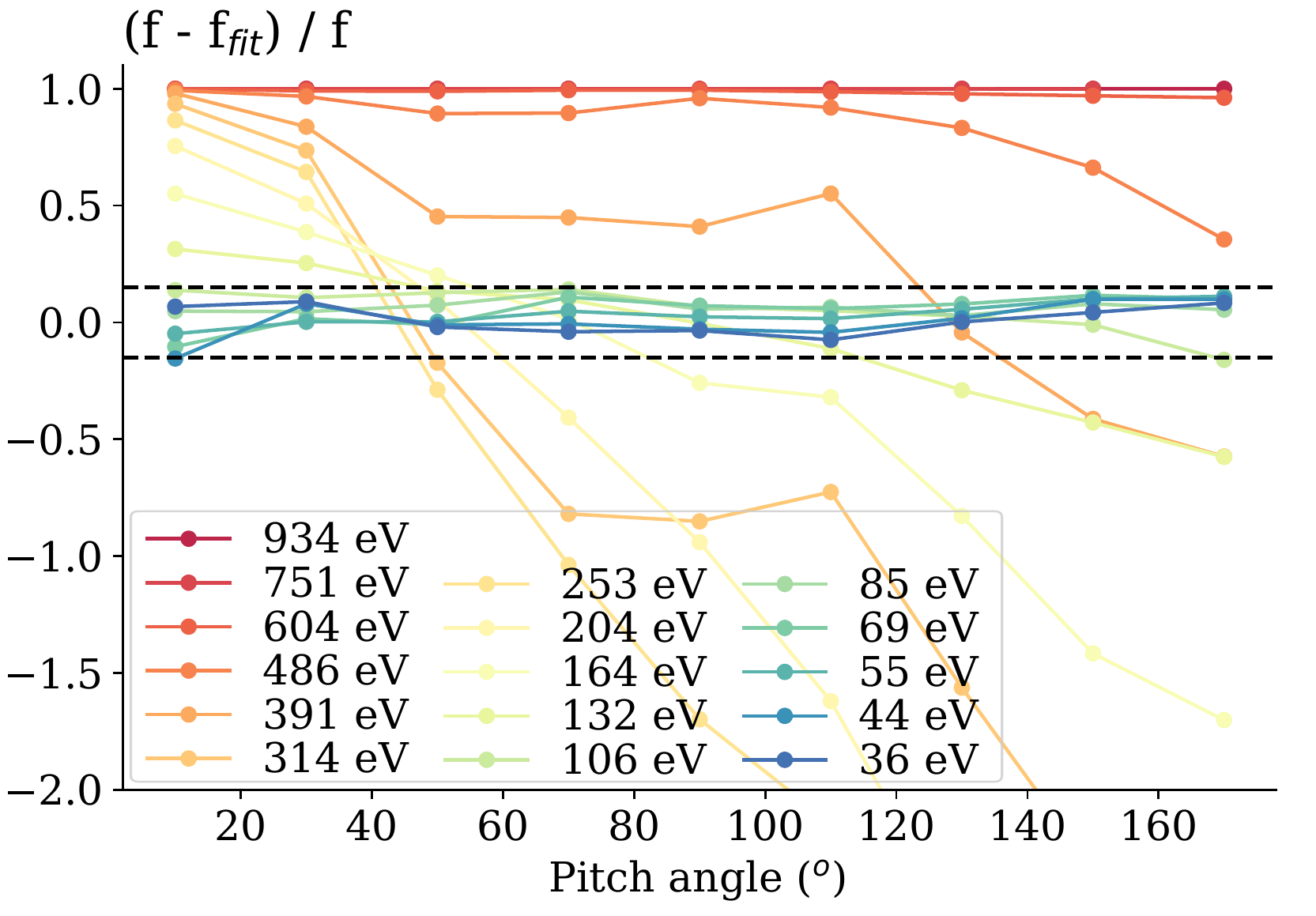}
    \caption{The normalised difference between the measured VDF and the core fit ($\Delta f_\mathrm{data,fit}$) binned in 20$^\circ$ pitch-angle bins. Different colours mark separate instrument energy bins. Black dashed line shows the criteria used in the determination of $E_\mathrm{C}$ and $E_\mathrm{BP}$.  }
    \label{fig:data_analysis}
\end{figure}

To define the boundaries between the deficit and the core, and between the core and the strahl, we look for departures of the measured VDF from the fitted bi-Maxwellian core distribution. We calculate the normalised difference between the two in each measured point as
\begin{equation}
    \Delta f_\mathrm{data,fit} = \frac{f_\mathrm{data} - f_\mathrm{fit}}{f_\mathrm{data}}, 
\end{equation}
where $f_\mathrm{data}$ are the measured values and $f_\mathrm{fit}$ are the core fit values corresponding to the centres of the measurement bins in phase space. We bin $\Delta f_\mathrm{data,fit}$ into 20$^\circ$-wide pitch-angle bins and calculate the median value in each bin ($\mathrm{med}\left\{\Delta f_\mathrm{data,fit} \right\}$). This way we avoid the possible effects of the FOV blockage by the heat shield, which is $\sim$ 10$^\circ$ wide in pitch angle, when magnetic field is aligned with radial direction \citep{Kasper2016}.  
Figure \ref{fig:data_analysis} shows these values for separate instrument energy bins on the example VDF from Figure \ref{fig:example_cut}. The value of $\Delta f_\mathrm{data,fit}$ at low energies, represented by the blue part of the colour-scale spectra, remains around 0, which means that the bi-Maxwellian fit represents well the electron VDF in this energy range. High energies are plotted in red and reach 1, which indicates that the measured VDF is much greater than the obtained core fit. The departure from the core fit at high energies is expected due to presence of the halo population. For the energies in between we observe a pitch-angle dependent evolution of the departures from the bi-Maxwellian core.

We define the strahl break-point energy ($E_\mathrm{BP}$) as the lowest energy at which $\mathrm{med}\left\{\Delta f_\mathrm{data,fit} \right\}$ exceeds the value of $0.15$ in the 0 - 20$^\circ$ pitch-angle bin. If this energy is greater than 40\,eV and less than 700\,eV, we consider it a successful determination of $E_\mathrm{BP}$. 

We define the electron cutoff energy ($E_\mathrm{C}$) as the lowest energy at which $\mathrm{med}\left\{\Delta f_\mathrm{data,fit} \right\}$ decreases below the value of $-0.15$ in the 160 - 180$^\circ$ pitch-angle bin. Black dashed lines mark the limit values in Figure \ref{fig:data_analysis}. The cutoff is only considered for instances when $\mathrm{med}\left\{\Delta f_\mathrm{data,fit} \right\}$ at any energy is less than $-0.5$. We move $E_\mathrm{C}$ to the Sun rest frame using the solar wind velocity.
The method is considered successful if the identified energy lies within an interval from 60\,eV to 400\,eV.

We calculate $\lambda_\mathrm{mfp}$ used in Eq. \ref{eq:ED} from the ratio between the electron parallel core thermal velocity ($w_\mathrm{c\parallel}$) and the electron collision frequency, which we obtain from the relation between the electron density and temperature \citep[e.g.][]{Salem2003}
\begin{equation}
\nu_\mathrm{e} = 2.9 \cdot 10^{-6} n_\mathrm{c} T_\mathrm{c\parallel}^{-3/2} \ln \Lambda,
\label{eq:nu}
\end{equation}
where $\ln \Lambda$ is the Coulomb logarithm defined as
\begin{equation}
\ln \Lambda = \ln \left( \frac{12 \pi (\epsilon_0 k_\mathrm{B} T_\mathrm{c\parallel})^{3/2}}{n_\mathrm{c}^{1/2}e^3} \right), 
\label{eq:lnL}
\end{equation}
 $\epsilon_0$ is the vacuum permittivity.

\section{Results} \label{sec:results}

\begin{figure*}
    \plotone{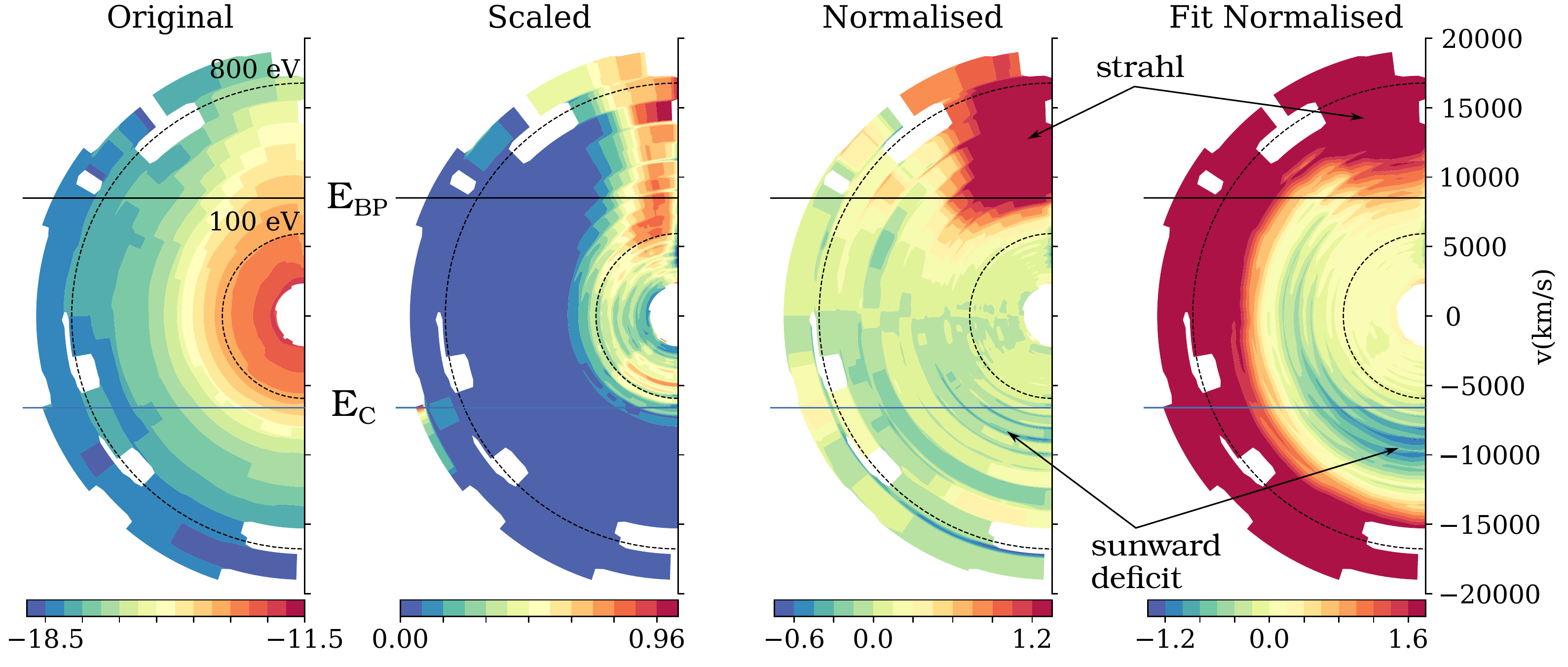}
    \caption{The same electron VDF as shown in Figures \ref{fig:example_cut} and \ref{fig:example_vdf}, plotted as a function of  parallel and perpendicular velocity. The VDF is shown in the magnetic field aligned frame, centred on the core parallel velocity. Plots from left to right present: the \emph{original} VDF; the \emph{scaled} VDF, where values in each energy bin are scaled between 0 and 1; the \emph{normalised} VDF, where the original VDF is divided by the perpendicular VDF cut ($f(v_\perp,v_\parallel=0)$); and the \emph{fit normalised} VDF, where the VDF is divided by the core electron fit.}
    \label{fig:example_vdf}
\end{figure*}

The Figure \ref{fig:example_vdf} shows the same electron VDF as shown in Figure \ref{fig:example_cut} but plotted against $v_\parallel$ and $v_\perp$. We show a single VDF using four different 2D representations, which we obtain by the integration along the angle perpendicular to the magnetic field. All plots show the distribution in the magnetic-field-aligned frame centred on the core parallel velocity. The representation marked \emph{original} shows the measured VDF values with a logarithmic colour-scale. In  the representation marked \emph{scaled}, each energy bin -- i.e., each circular belt in $(v_\parallel, v_\perp)$ parameter space -- is scaled to a value between 0 and 1, where 1 corresponds to the maximum value of the VDF in the given energy belt. This representation removes the information about the absolute value of the VDF and its strong gradient in energy. The benefit of the \emph{scaled} VDF is the exposure of smaller anisotropic features at all energies. In cases for which two features arise in the same energy bin, the scaled VDFs can be misleading though as they highlight only the stronger feature. The representation marked \emph{normalised} is obtained by dividing each of the VDF values with  $f(v_\perp,v_\parallel=0)$ of the associated energy bin. Pitch-angle directions in which the distribution function is less than $f(v_\perp,v_\parallel=0)$, appear in green, and those in which the distribution function is greater than $f(v_\perp,v_\parallel=0)$ appear in red. The  representation marked \emph{fit-normalised} shows the logarithm of electron VDF divided by the core fit. Yellow phase space regions are well represented by a bi-Maxwellian distribution while departures are seen in red and blue colours. 

We use these representations to obtain a better understanding of the 2D shape of the features in the electron VDF.
Electrons at energies below $\sim$ 100\,eV are predominantly members of the almost isotropic core population, which shows almost no pitch-angle variation in all representations. At positive $v_\parallel$, we observe a distinct strahl, the shape of which is most clearly defined in the \emph{scaled} VDF. Its width in terms of perpendicular velocity appears almost constant, which gives a decreasing angular pitch-angle width with increasing electron energy. The sunward deficit is present at negative $v_\parallel$, and shows in green and blue colours in the \emph{fit-normalised} VDF. This feature not only persists at high pitch angles (close to 180$^\circ$), but it forms a circular belt in phase space at some energies reaching to the strahl population at small pitch angles. The absence of the deficit in the \emph{normalised} VDF tells us that this feature is close to isotropic, existing also at pitch angles around 90$^\circ$. \\

\begin{figure}
    \plotone{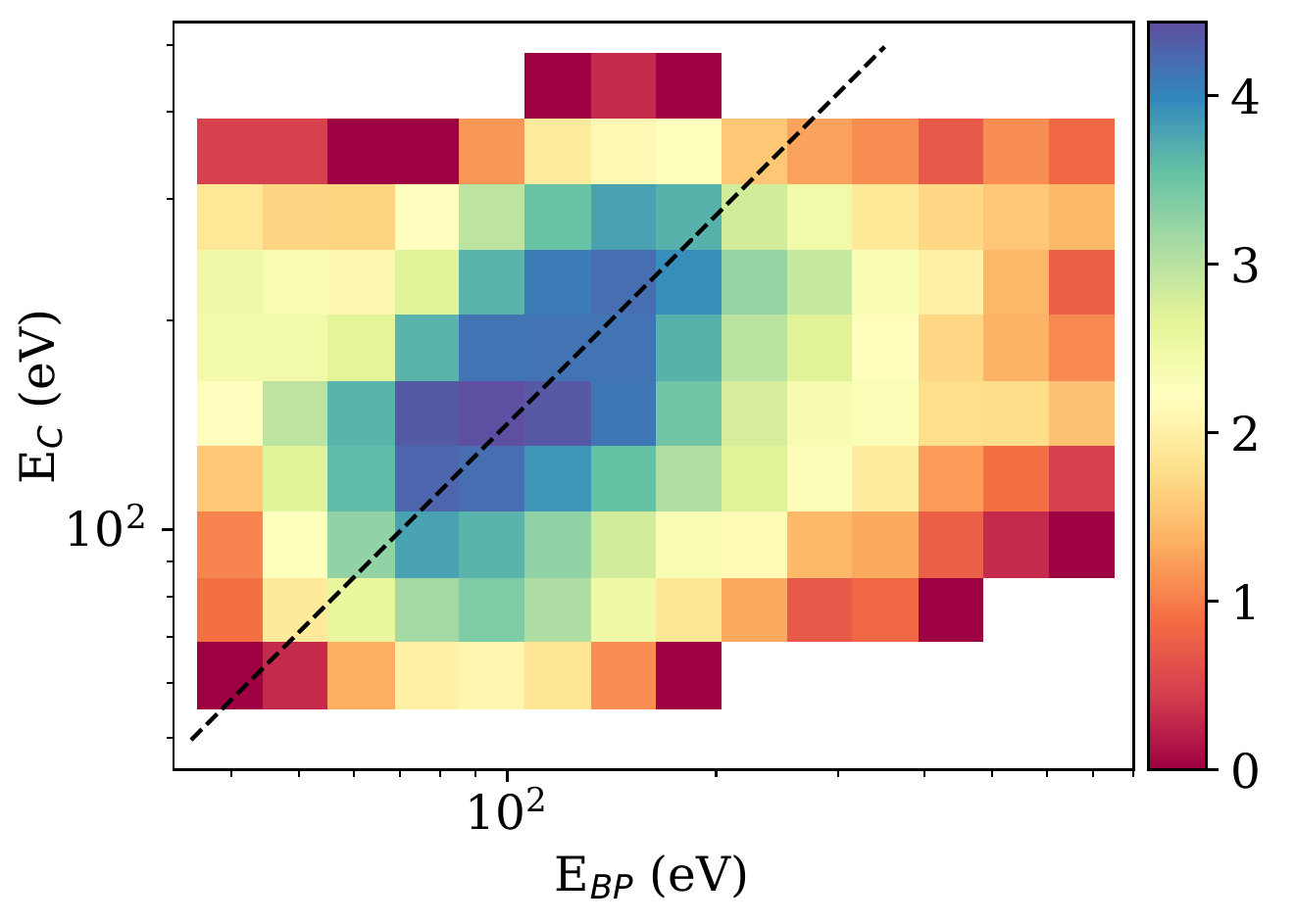}
    \caption{A 2D histogram showing the relation between $E_\mathrm{BP}$ and $E_C$. The bin size corresponds to the energy resolution of the SPAN-E instrument. The colour-scale represents the logarithm of the number of instances in each bin ($\log$(\#)). The dashed line denotes $E_\mathrm{C} = 1.42 E_\mathrm{BP}$.}
    \label{fig:EBP_EC}
\end{figure}

In total, we successfully fit 510,610 full 3D electron VDFs, out of which we determine $E_\mathrm{BP}$ in 98.3 \% and $E_\mathrm{C}$ in 55.4 \% of the cases. We statistically visualise the energy at which these two boundaries occur in a histogram in Figure \ref{fig:EBP_EC}, where the bin sizes correspond to the instrument's energy resolution. $E_\mathrm{BP}$ and $E_\mathrm{C}$ are strongly correlated and the mean ratio between the two ($E_\mathrm{C}/E_\mathrm{BP}$) equals to 1.42.

\begin{figure*}
    \plotone{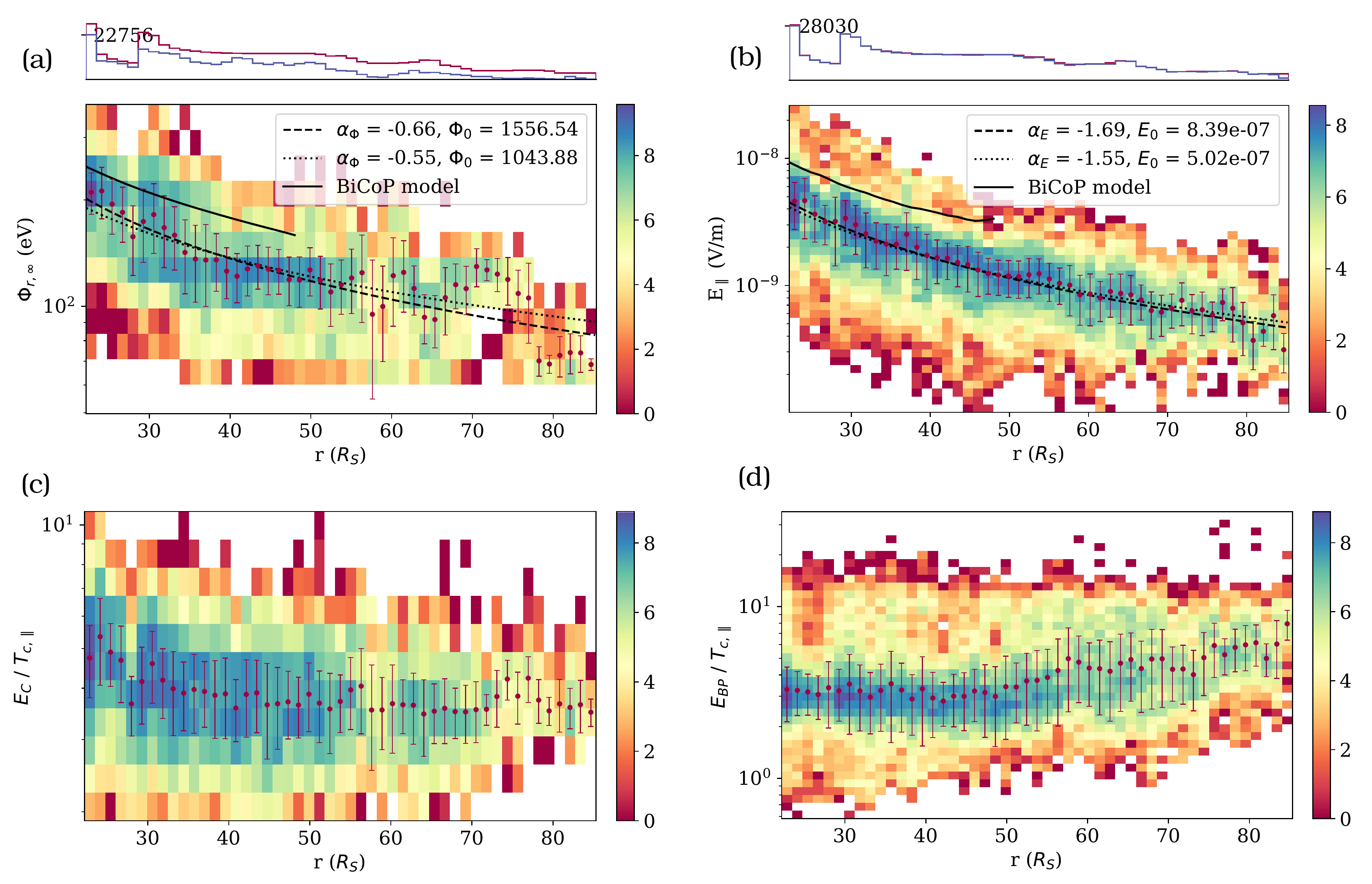}
    \caption{A 2D histogram showing the radial evolution of (a) $\Phi_\mathrm{r,\infty}$, (b) $\mathrm{E}_\parallel$, (c) the ratio between $E_\mathrm{C}$ and $T_\mathrm{c\parallel}$, and (d) the ratio between $E_\mathrm{BP}$ and $T_\mathrm{c\parallel}$. The colour-scale represents the logarithm of the number of instances in each bin ($log$(\#)). The red dots represent the median of samples in each radial bin with an error bar showing one standard deviation from the mean value in each bin. The dashed line represents the best power law fit (Eq. \ref{eq:power_law_phi} and \ref{eq:power_law_E}) to the data points, with the fitted parameters marked in the legend. The dotted line represents a fit where the power law index was fixed to equal the results from BiCoP \citep{Bercic2021}. Above the main plot, 1D histograms show the number of data points in each of the radial bins. The number of all  available data points is shown in red and the number of the data points used in the 2D histogram in blue.  }
    \label{fig:Phi_E}
\end{figure*}

We plot $\Phi_\mathrm{r,\infty}$ obtained from $E_\mathrm{C}$ through Eq. \ref{eq:phi} on a 2D histogram against radial distance in Figure \ref{fig:Phi_E}a. The red dots mark the median values, and the corresponding error bars show the intervals of one standard deviation of the data set in each radial bin. $\Phi_\mathrm{r,\infty}$ decreases with radial distance and takes the values between 300 and 60\,V. The step histograms above the plot compare the number of available data points in each radial bin to the number of data points for which $E_\mathrm{C}$ is found. According to these histograms, the proportion of the electron VDFs with sunward deficit decreases with increasing radial distance. 

We fit all data points with a power law distribution of the form
\begin{equation}
    \Phi_\mathrm{r,\infty} = \Phi_0 r^{\alpha_\Phi},
    \label{eq:power_law_phi}
\end{equation}
where $\Phi_0$ is a constant and $\alpha_\Phi$ is the power law index. The best fit is plotted with the black dashed line in Figure \ref{fig:Phi_E}a and the fitting parameters noted in the legend. The black dotted line shows a fit with $\alpha_\Phi$ fixed to the value obtained from the radial evolution of the ambipolar potential in BiCoP simulations \citep{Bercic2021}. 

We show a similar plot, but for $\mathrm{E}_\parallel$ calculated from $E_\mathrm{BP}$ through Eq. \ref{eq:E} in Figure \ref{fig:Phi_E}b. The absolute values of $\mathrm{E}_\parallel$ span between 0.5 to 10\,nV/m and decrease faster with radial distance than $\Phi_\mathrm{r,\infty}$. We fit the data points with a power law
\begin{equation}
    \mathrm{E}_\parallel = \mathrm{E}_0 r^{\alpha_\mathrm{E}},
    \label{eq:power_law_E}
\end{equation}
and mark the values of the fitting parameters $\mathrm{E}_0$ and $\alpha_\mathrm{E}$ in the legend. As for the case of the ambipolar potential, we show a second fit as a dotted line, representing a power law function with an index equal to the one obtained from the BiCoP model.

In the bottom row of Figure \ref{fig:Phi_E}, we explore how the boundary energies $E_\mathrm{C}$ and $E_\mathrm{BP}$ compare to $T_\mathrm{c\parallel}$ for different radial distances. The ratio $E_\mathrm{C}/T_\mathrm{c\parallel}$ slightly decreases with radial distance, spanning from a mean value of 5.7 close to the Sun, to 3.3 farther away. The ratio $E_\mathrm{BP}/T_\mathrm{c\parallel}$ exhibits an opposite trend, increasing from the mean value of 3.0 to 7.8.

\section{Discussion} \label{sec:discussion}
\subsection{Ambipolar electric potential ($\Phi_\mathrm{r,\infty}$)}

The electron deficit in the suprathermal energy range is reported already by \citet{Pilipp1987b}, analysing Helios data which covers distances down to 65\,$R_S$ from the Sun. In the PSP data, the sunward deficit is a common feature \citep{Halekas2019,Bercic2020}, which contributes significantly to the net electron heat-flux \citep{Halekas2020arxiv}. The characteristics of the sunward deficit and their relation to different solar wind parameters are investigated by \citet{Halekas2021submitted}.

In the collisionless exospheric models \citep{Lemaire1970, LemaireJosephandScherer1971, Jockers1970, Maksimovic1997c, Pierrard1999, Maksimovic2001a, Zouganelis2004a} no electrons with energies greater than the electric potential energy ($\mathcal{E}_\Phi$) exist in the sunward portion of the electron VDF. Therefore, knowing the electron cutoff energy, we can obtain the value of the electric potential at any radial distance in the exosphere (Eq. \ref{eq:phi}). In kinetic models accounting additionally for Coulomb collisions, the electron cutoff is smoothed, appearing more like a deficit compared to the expected Maxwellian core VDF \citep{pierrard2001, Bercic2021}. These models reproduce the observed radial profiles of the electron core properties, like density, temperature and anisotropy. They also reproduce the strahl; however, they fail to produce the halo population. Electron VDFs observed close to the Sun (see example in Figures \ref{fig:example_cut} and \ref{fig:example_vdf}) exhibit only a tenuous halo population and are thus very similar to the VDFs from BiCoP simulation \citep[see Figures 7 and 8 by][]{Bercic2021}. Comparing the \emph{normalised} VDF in Figure \ref{fig:example_vdf} with the \emph{normalised} VDF in Figure 7 in \citet{Bercic2021}, we see however, that the shape of the sunward deficit is slightly different. In PSP data the deficit exists at pitch angles $\gtrsim 45^\circ$, while in BiCoP it only takes the angles $\gtrsim 135^\circ$.

In the near-Sun solar wind (at $\sim$ 34\,$R_S$), Coulomb collisions only effectively scatter the strahl electrons with energies smaller than 250\,eV \citep{Horaites2018a, BoldyrevHoraites2019, Bercic2021}. The scattering of the strahl at higher energies and the creation of the halo population are therefore attributed to other phenomena, including wave--particle interactions \citep{Vocks2005a, Kajdic2016a, Verscharen2019scat,Jeong2020, KrishnaJagaramuldi2020, Catell2021arXiv} and scattering by  background turbulence \citep{Pagel2007a, Saito2007a}. Observational studies by \citet{Stverak2009a} and \citet{Halekas2019} suggest that the halo is more prominent farther from the Sun, which could be the reason why the sunward deficit has not been observed at larger radial distances (Figure \ref{fig:Phi_E}a). 

A recent study by \citet{Bercic2021wh}, who analyse Solar Orbiter \citep{Muller2020} in situ measurements of solar-wind electrons, shows that the sunward deficit can drive the growth of quasi-parallel whistler waves, leading to quasi-linear electron diffusion in velocity space. This new proposed instability tends to fill the sunward deficit and may thus also be the reason why the electron cutoff ceases to exist at larger radial distances.

Another possibility for the disappearance of the deficit could simply be the Coulomb collisions: as $\Phi_\mathrm{r,\infty}$ decreases with radial distance it moves to the energy range where electron collisions are frequent. They could completely erase the residue of the cutoff. 

Ballistic electrons, in exospheric models representing the electron core population, are limited in energy to a range $\lesssim \mathcal{E}_\Phi$. Therefore, we expect that the core electron temperature ($T_\mathrm{c}$) follows the radial evolution of $\Phi_\mathrm{r,\infty}$. This would show as a constant ratio between $E_\mathrm{C}$ and $T_\mathrm{c \parallel}$ in Figure \ref{fig:Phi_E}c. The observed variation in the ratio is not large, but a slight decreasing trend is present, mostly for radial distances below 35\,$R_S$. The increase in  $E_\mathrm{C}/T_\mathrm{c \parallel}$ with decreasing heliocentric distance suggests the increasing importance of Coulomb collisions, which smear the exospheric VDF features and raise $E_\mathrm{C}$.

We compare the measured $\Phi_\mathrm{r,\infty}$ to the results of a kinetic numerical model (BiCoP) \citep{Bercic2021}, which builds up a supersonic radially expanding solar wind taking into account binary particle collisions and the self-consistent $\mathrm{E}_\parallel$ \citep{LandiPantellini2001, Landi2003a}.
The simulation box spans from 3 to 49\,$R_S$, thus overlapping with approximately half of the radial interval shown in this study. The self-consistently obtained $\Phi_\mathrm{r,\infty}$ from BiCoP is added to Figure \ref{fig:Phi_E}a and evolves with radial distance as a power law with an index $\alpha_\mathrm{\Phi, BCP}= -0.55$. This result is close to the power law index obtained in our observational study, $\alpha_\mathrm{\Phi}= -0.66$. In Figure \ref{fig:Phi_E}a, we show a second fit to the data points with fixed $\alpha_\mathrm{\Phi}= -0.55$ (dotted line) to emphasise that the power law index is a sensitive fitting coefficient, and may vary across different solar wind streams. For the scope of this work, we only calculate the average properties of all of the solar wind measured during the four PSP encounters.

Our experimentally determined $\alpha_\mathrm{\Phi}$ diverges from the power law index in collisionless exospheric models, $\alpha_\mathrm{\Phi, Exo} = -1.33$ \citep[e.g.][]{MeyerVernet1998, Zouganelis2004a}, indicating that collisions play an important role in the radial evolution of $\Phi_\mathrm{r,\infty}$ and in the ambipolar solar wind acceleration. An analytical solution of the drift-kinetic equation including the effects of Coulomb collisions \citep{Boldyrev2020} gives a power law with $\alpha_\mathrm{\Phi, DK} = -0.4$, which is closer to our observations.

We add $\Phi_\mathrm{r,\infty}$ obtained with a medium-collisional BiCoP run \citep[MC in][]{Bercic2021} to Figure \ref{fig:Phi_E}a. The modelled values are within the range of the observed values, even though the BiCoP boundary parameters -- electron and proton temperature at 3\,$R_S$ set to 121\,eV --  differ from the expected coronal temperatures. Electrons in the corona are observed to be colder, $T_\mathrm{e} \sim$ 86\,eV (1\,MK) \citep{Cranmer2002,Bercic2020, Stansby2020}, while the proton temperature is expected to be greater. This difference in temperature between the two species and the preferential perpendicular heating of solar wind protons potentially result in the observed $\Phi_\mathrm{r,\infty}$, even when the electron temperature at the origin is less than the electron temperature assumed in BiCoP simulations. Initialising BiCoP runs with different coronal temperatures for electrons and protons, and with varying proton anisotropies would give further insight into this phenomenon. 

\subsection{Ambipolar electric field ($\mathrm{E}_\parallel$)}

The SERM \citep{Scudder1996, Scudder2019serm, Scudder2019crisis, Scudder2019astro} accounts for the effects of the global ambipolar electric field ($\mathrm{E}_\parallel$) in the presence of Coulomb collisions. The Dreicer electric field ($E_\mathrm D$, Eq. \ref{eq:ED}) serves as a measure of the importance of these two competing phenomena \citep{Dreicer1959, Dreicer1960}. We use the measured $E_\mathrm{BP}$ to estimate the ambipolar electric field in the solar wind (Eq. \ref{eq:E}). Before discussing the properties of the observed $\mathrm{E}_\parallel$, we compare $E_\mathrm{BP}$ in the near-Sun solar wind to already existing studies.

The idea that the separatrix between the thermal and suprathermal electron populations contains information about the electron kinetics is discussed in early observational studies \citep{Feldman1975a, Pilipp1987b}. \citet{ScudderOlbert1979a} theoretically predict that the break-point energy scales with the local electron temperature as $E_\mathrm{BP} = 7 k_\mathrm B T_\mathrm c$. This value agrees with the break-point between the core and the halo obtained by \citet{Stverak2009a}, who analyse electron VDFs from Helios, Cluster, and Ulysses. However, the ratio $E_\mathrm{BP}/T_\mathrm{c\parallel}$ corresponding to the strahl population assumes slightly lower values, between 2 (at 0.3\,au) and 5 (at 3\,au). Similar values are obtained by \citet{Landi2012b} using a kinetic BiCoP simulation. At radial distances between 1 and 3\,au, they find that $E_\mathrm{BP}/T_\mathrm{c\parallel}$ varies between 1 and 4 and depends mainly on the density of the modelled solar wind. In a more recent study, including Cluster data, \citet{Bakrania2020} obtain the ratio of 5.5 at 1 au as well as an anti-correlation between the strahl-$E_{BP}$ and the solar wind velocity.

The ratio $E_\mathrm{BP}/T_\mathrm{c\parallel}$ obtained from the PSP data shown in our work (Figure \ref{fig:Phi_E}d) agrees with previous Helios results. Its median value is approximately constant, $\sim$ 3, up to a radial distance of 50\,$R_S$. At greater distances, it approaches $\sim$ 6.

For the majority of the samples, we find $E_\mathrm{BP}<E_\mathrm{C}$, which suggests that electrons with energy less than $E_\mathrm{C}$ are not limited to Maxwellian core electrons but include a small part of strahl electrons. The same is seen in the BiCoP kinetic solar wind model \citep{Bercic2021}.

We present the first observational estimates of $\mathrm{E}_\parallel$ in the solar wind (Fig. \ref{fig:Phi_E}b). Its strength is of order a few \,nV/m, and, as expected, decreases with radial distance. The radial evolution is best represented by a power law function with an index $\alpha_\mathrm{E} = -1.69$. This index is very close to the power index resulting from BiCoP simulations, $\alpha_\mathrm{E, BCP}= -1.55$ \citep{Bercic2021}. Following the same method as for $\Phi_\mathrm{r,\infty}$, the dotted curve in Figure \ref{fig:Phi_E}b shows a power law fit to the data points with a fixed index of $-1.55$. For comparison, we plot $\mathrm{E}_\parallel$ from the medium-collisional BiCoP run to Figure \ref{fig:Phi_E}b as a black line.

The parameters $\Phi_\mathrm{r,\infty}$ and $\mathrm{E}_\parallel$ are related to each other as
\begin{equation}
    \Phi_\mathrm{r,\infty} = \int_r^\infty \mathrm{E}_\parallel (r) \,dr.
    \label{eq:phi-E}
\end{equation}
If we assume that $\Phi_\mathrm{r,\infty}$ and $\mathrm{E}_\parallel$ follow power laws in $r$, then the difference between the power law indices of each of the quantities should be equal to 1 ($\alpha_\mathrm{E} - \alpha_\mathrm{\Phi} = -1$). Our results agree with this theoretical relation within the measurement uncertainty, as the difference between the fitted power law indices is 1.02.

\begin{figure*}
    \plotone{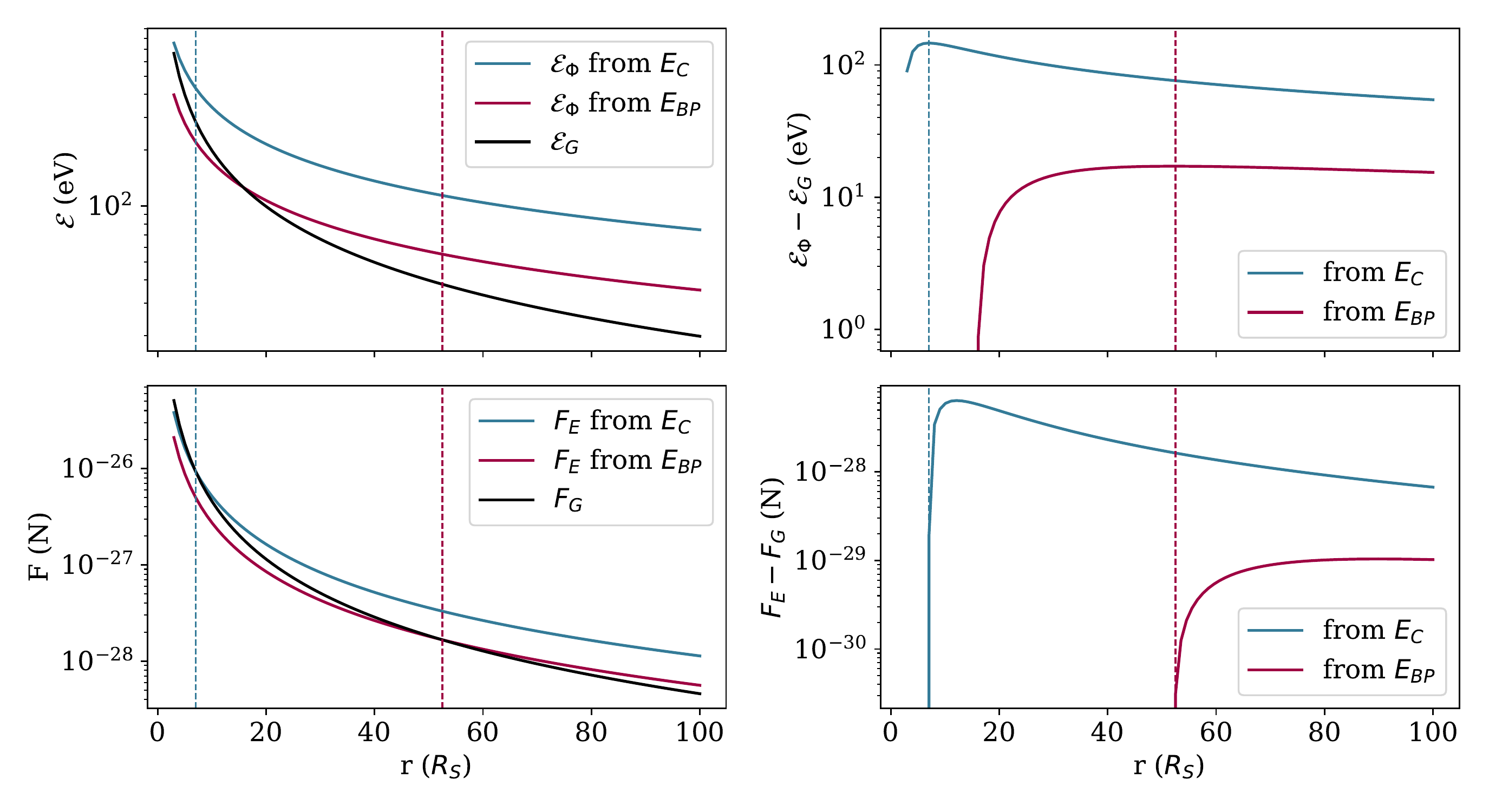}
    \caption{(a) Radial evolution of the proton gravitational energy ($\mathcal{E}_G$) and the electric potential energy ($\mathcal{E}_\Phi$); (b) Solar wind proton energy balance; (c) Radial evolution of the gravitational force ($F_G$) and the electric force ($F_E$) for a proton; (d) The resulting net force on a proton. In all panels, blue colour corresponds to the solution obtained from $E_\mathrm{C}$, and red colour to the solution obtained from $E_\mathrm{BP}$. Vertical dashed lines mark $r_\mathrm{max}$.}
    \label{fig:E-F}
\end{figure*}

\subsection{Ambipolar contribution to the acceleration of the solar wind}

\begin{figure}
    \plotone{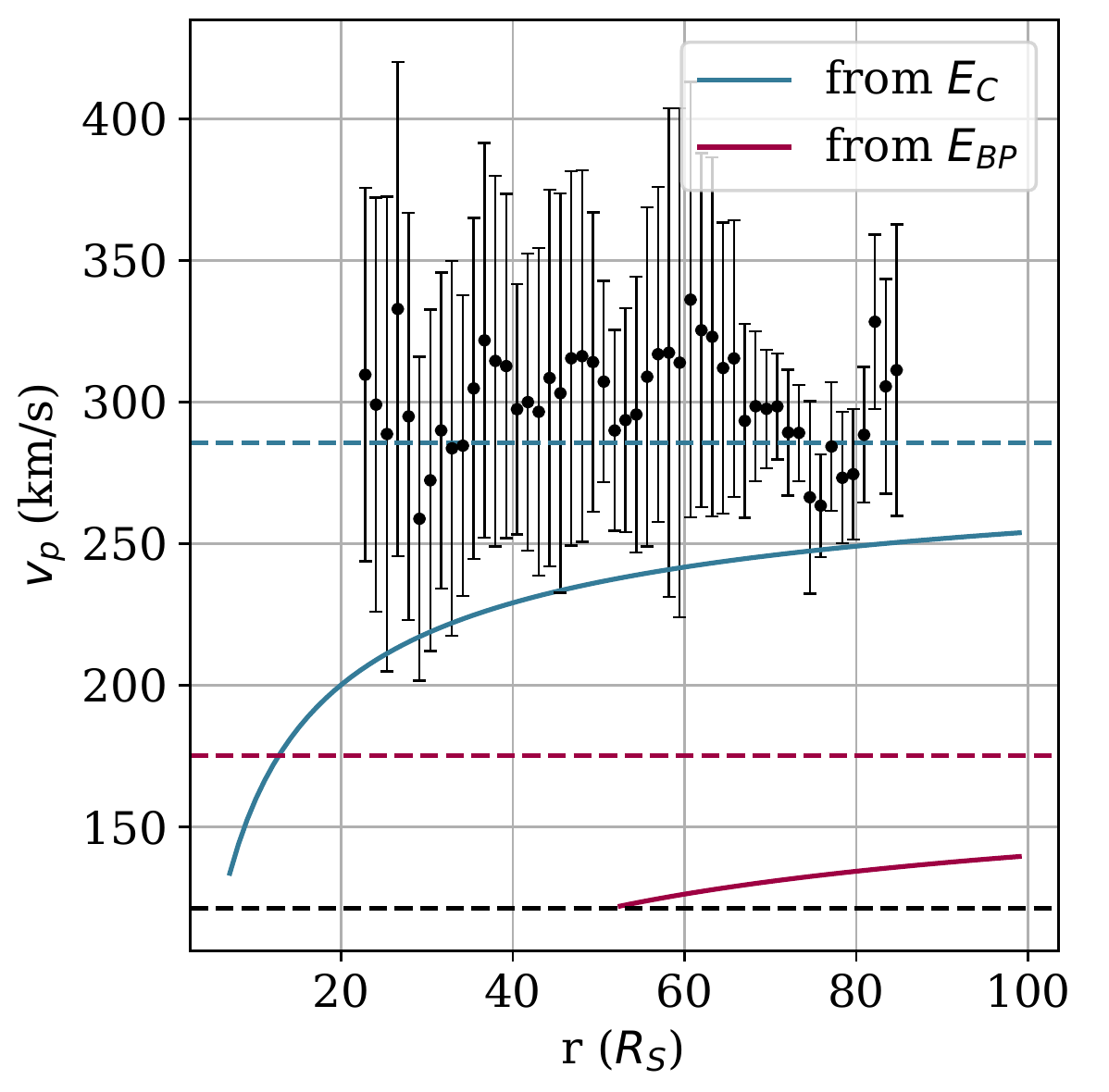}
    \caption{Radial evolution of the solar wind proton velocity ($v_\mathrm{p}(r)$) obtained via Eq. \ref{eq:vel}. The blue and pink lines mark the solutions resulting from different models and different features of the electron VDFs. The dashed blue and pink lines mark their asymptotic values. The black dashed line marks the bulk velocity $v_\mathrm{p}(r_\mathrm{max})$ corresponding to $T_\mathrm{p\parallel} = 0.7$\,MK. The black dots with belonging error bars show the mean value and the standard deviation of the observed solar wind velocity binned along the radial direction in the same way as in Figure \ref{fig:Phi_E}. }
    \label{fig:v_sw}
\end{figure}

The total solar wind proton potential energy $\Psi(r)$ is the sum of the repulsive electric potential energy $\mathcal{E}_\Phi (r)$ and the attractive gravitational potential energy $\mathcal{E}_G (r)$. We use the fitted curve from Figure \ref{fig:Phi_E}a to  calculate $\mathcal{E}_\Phi(r)$ from $\Phi_\mathrm{r,\infty}$ as
\begin{equation}
    \mathcal{E}_\Phi (r) = e\Phi_0 r^{\alpha_\Phi}
\end{equation}
based on our determination of $E_C$.
Likewise, we use the fitted curve from Figure \ref{fig:Phi_E}b to  calculate $\mathcal{E}_\Phi(r)$ from $\mathrm{E}_\parallel$ as
\begin{equation}
    \mathcal{E}_\Phi (r) =  -\frac{e\mathrm{E}_0}{r_0(\alpha_\mathrm{E}+1)} r^{\alpha_E+1}
\end{equation}
based on our determination of $E_\mathrm{BP}$.
The gravitational energy of a proton is defined as
\begin{equation}
    \mathcal{E}_G (r) = \frac{GM_\mathrm{S}}{r}m_\mathrm{p}, 
\end{equation}
where $G$ is the gravitational constant, $M_\mathrm{S}$ the mass of the Sun and $m_\mathrm{p}$ the mass of a proton. The results are shown in the top row of Figure \ref{fig:E-F}. $\mathcal{E}_G$ decreases with radial distance faster than $\mathcal{E}_\Phi$, which means that $\mathcal{E}_\Phi$ dominates at larger radial distances, and $\Psi(r)$ peaks at a radial distance, which we denote $r_\mathrm{max}$. The kinetic theory \citep{Scudder1996} and BiCoP numerical simulations \citep{Landi2003a} predict a maximum of $\Psi(r)$ near the proton sonic point.

The bottom row of Figure \ref{fig:E-F} shows the radial evolution of the energy gradients, corresponding to the electric force $F_\mathrm{E}$ and the gravitational force $F_\mathrm{G}$. At small radial distances, $F_\mathrm{G}> F_\mathrm{E}$, so that the net force on protons points towards the Sun. The radial distance at which $F_\mathrm{G} = F_\mathrm{E}$ is marked with a dashed line and corresponds to the location of the total energy peak, $r_\mathrm{max}$. All protons with $v_\parallel\geq0$ present at $r_\mathrm{max}$ can escape the Sun's gravitational potential, as the net force on them above this distance is only positive. We calculate the radial evolution of the velocity gained by a test proton ($v_\mathrm{p\Psi}(r)$) moving in the total potential $\Psi(r)$ through integration of the net force, $F(r) = F_\mathrm{E} (r) - F_\mathrm{G} (r)$, above $r_\mathrm{max}$ as
\begin{equation}
    v_\mathrm{p\Psi} (r) = \sqrt{2 \int_{r_\mathrm{max}}^r \frac{F(r)}{m_\mathrm{p}}dr}.
\end{equation}

The terminal test proton velocities $v_\mathrm{p\Psi}(\infty)$ result in 164\,km\,s$^{-1}$ from the method using $E_\mathrm{C}$, and 54\,km\,s$^{-1}$ from the method using $E_\mathrm{BP}$. 

To calculate the bulk velocity of the protons $v_\mathrm{p}(r)$, we follow the exospheric approach. We assume a Maxwellian proton distribution at $r_\mathrm{max}$, with a parallel temperature $T_\mathrm{p\parallel} = 0.7$\,MK. This is an estimation of the $T_\mathrm{p\parallel}$ at 7\,$R_S$ following from the extrapolation of the radial trends presented by \citet{Maksimovic2020}. The proton parallel temperature in this simple approach does not vary with radial distance, so $v_\mathrm{p}(r)$ is
\begin{equation}
    v_\mathrm{p}(r) = v_\mathrm{p\Psi} (r) + v_\mathrm{p}(r_\mathrm{max}), 
     \label{eq:vel}
\end{equation}

where 
\begin{equation}
    v_\mathrm{p}(r_\mathrm{max}) = \frac{2w_\mathrm{p\parallel}}{\sqrt{\pi}}.
\end{equation}

$w_\mathrm{p\parallel}$ is the proton parallel thermal velocity defined as

\begin{equation}
    w_\mathrm{p\parallel} = \sqrt{\frac{2k_\mathrm B T_\mathrm{p\parallel}}{m_\mathrm{p}}}.
\end{equation}

For simplicity we use the same $v_\mathrm{p}(r_\mathrm{max}) = 121$\,km\,s$^{-1}$ for both obtained solutions, even though they exhibit different $r_\mathrm{max}$.

Figure \ref{fig:v_sw} shows the obtained velocity curves together with their asymptotic values, marked with blue and pink dashed lines. Black dashed line denotes $v_\mathrm{p}(r_\mathrm{max})$. $v_p (r)$ resulting from $E_\mathrm{C}$ is greater than $v_p(r)$ resulting from $E_\mathrm{BP}$, reaching a terminal velocity of 286\,km\,s$^{-1}$. At the radial distance of 45\,$R_S$ the average observed proton velocity is 303\,km\,s$^{-1}$. The resulting $v_p (45\,R_S) = 233$\,km\,s$^{-1}$ corresponds to 77\% of the observed velocity, or 59\% of the proton kinetic energy. This means that 23\% of the measured solar wind velocity must be gained through other solar wind acceleration mechanisms.

$v_p (r)$ obtained from $E_\mathrm{BP}$ related to the SERM model is smaller, with a terminal velocity of 175\,km\,s$^{-1}$. This result at first appears unphysical, since $r_\mathrm{max} = 53\,R_S$, which would suggest that below this distance we should not observe supersonic protons at all. However, depending on the location of the solar wind acceleration by mechanisms other than $\mathrm{E}_\parallel$, the contribution of the ambipolar acceleration could increase. Additional kinetic energy close to the Sun could produce a positive net force at smaller radial distances, creating more space for the ambipolar acceleration. Comparing the obtained terminal speed with the typical solar wind speed at 1\,au, we find that the  $\mathrm{E}_\parallel$ is responsible for 44\% of the solar wind velocity and 19\% of proton kinetic energy.

Note that this is only a crude estimation, as the model we use is simplified and includes strong assumptions. On the other hand, the statistical errors arising from the data analysis and the fits in Figures \ref{fig:Phi_E}a and \ref{fig:Phi_E}b are small. We do not include them in Figure \ref{fig:v_sw}, because this could be misleading for the reader.

Since this is the first effort to empirically quantify the acceleration of the solar wind by $\mathrm{E}_\parallel$, it is difficult to make conclusions on which of the separately obtained results is more valid. $\Phi_\mathrm{r,\infty}$ calculated from $E_\mathrm{C}$ is potentially overestimated, because we can not be sure that the energy associated with the sunward deficit directly corresponds to the electron cutoff in collisionless exospheric models. The boundary could be, as a consequence of Coulomb collisions, pushed towards higher energies. 

On the other hand, $\mathrm{E}_\parallel$ may be underestimated. In our analysis, we use the assumption that the separatrix between the overdamped and underdamped region is the same in the direction along with and opposite to the direction of the electric force on the electrons. We use this approximation because it appears consistent with the BiCoP simulations \citep{Bercic2021}, but theoretical work by \citet{Dreicer1960} and \citet{Fuchs1986} suggests that the boundary is asymmetric and appears at higher energy in the direction opposite to the electric force. This would in our analysis lead to a multiplication factor in Eq. \ref{eq:E} and consequentially higher terminal solar wind velocities. Further investigations are needed to relate the 2D shape of the separatrix in the observed VDFs to the 2D shape predicted by theoretical models.

\section{Conclusions}  \label{sec:concl}

We analyse electron VDFs measured by PSP in the near-Sun solar wind during its orbits 4 to 7. We identify the electron energies at which the measured distribution departs from the bi-Maxwellian core electron fit in the direction parallel and anti-parallel to the magnetic field. In the sunward part of phase space, we define the cutoff energy ($E_\mathrm{C}$) that marks the appearance of the sunward electron deficit. In the anti-sunward portion of phase space, we define the strahl break-point energy ($E_\mathrm{BP}$) that marks the start of the strahl population.
While the strahl is detected in almost all of the analysed distributions, the sunward deficit is only found in 56.8\% of the cases. The relative amount of electron VDFs with a sunward deficit decreases for larger radial distances.

We relate $E_\mathrm{C}$ to the electron cutoff in exospheric solar wind models, which allows us to estimate the ambipolar potential between the observation point and the asymptotic potential at large heliocentric distances ($\Phi_\mathrm{r,\infty}$). The resulting $\Phi_\mathrm{r,\infty}$ decreases with radial distance as $r^{-0.66}$. Its radial trend agrees with the results of the kinetic BiCoP model \citep{Bercic2021}, while its magnitude is slightly smaller than $\Phi_\mathrm{r,\infty}$ obtained numerically.

We assume that $E_\mathrm{BP}$ represents the separatrix between collisionally overdamped and underdamped regions of phase space, defined in the Steady Electron Runaway Model \citep{Scudder2019serm}. This allows us to estimate the ambipolar electric field in the solar wind ($\mathrm{E}_\parallel$). The estimated $\mathrm{E}_\parallel$ is of order 1\,nV/m and decreases with radial distance as $r^{-1.69}$.

We finally calculate the total proton potential energy $\Psi(r)$, separately from $\Phi_\mathrm{r,\infty}$ and $\mathrm{E}_\parallel$ to estimate the contribution of the ambipolar acceleration to the total solar wind acceleration. From the approach following the exospheric models and $E_\mathrm{C}$ we find a terminal solar wind velocity 286\,km\,s$^{-1}$. Following the SERM model and $E_\mathrm{BP}$ we find a terminal solar wind velocity 175\,km\,s$^{-1}$. In the first case we are able to directly compare the observed solar wind velocity with the calculated ambipolar contribution, which amounts to 77\% at the radial distance of 45\,$R_S$.

\begin{acknowledgments}
The SWEAP and FIELDS experiments on the Parker Solar Probe spacecraft were designed and developed under NASA contract NNN06AA01C.
L.~B., C.~J.~O., and D.~V. are supported by STFC Consolidated Grant ST/S000240/1. D.~V.~is supported by STFC Ernest Rutherford Fellowship ST/P003826/1.
\end{acknowledgments}

\bibliography{manuscript}{}
\bibliographystyle{aasjournal}



\end{document}